\documentclass[aps,prc,floatfix,twocolumn,superscriptaddress]{revtex4}
\usepackage{epsf}
\usepackage{amsmath,amssymb}
\usepackage{graphicx}
\usepackage{wrapfig}
\usepackage{longtable}
\usepackage{array}
\usepackage{color}

\def\deg{^\circ}

\begin{document}

\title{Direct observation of quark-hadron duality in the free neutron
	$F_2$ structure function}

\author{I.~Niculescu}
\affiliation{James Madison University, Harrisonburg, Virginia 22807, USA}
\author{G.~Niculescu}
\affiliation{James Madison University, Harrisonburg, Virginia 22807, USA}
\author{W.~Melnitchouk}
\affiliation{Thomas Jefferson National Accelerator Facility, Newport News, Virginia 23606, USA}
\author{J.~Arrington}
\affiliation{Argonne National Laboratory, Argonne, Illinois 60439, USA}
\author{M.~E.~Christy}
\affiliation{Hampton University, Hampton, Virginia 23668, USA}
\author{R.~Ent}
\affiliation{Thomas Jefferson National Accelerator Facility, Newport News, Virginia 23606, USA}
\author{K.~A.~Griffioen}
\affiliation{College of William and Mary, Williamsburg, Virginia 23187, USA}
\author{N.~Kalantarians}
\affiliation{Hampton University, Hampton, Virginia 23668, USA}
\author{C.~E.~Keppel}
\affiliation{Thomas Jefferson National Accelerator Facility, Newport News, Virginia 23606, USA}
\author{S.~Kuhn}
\affiliation{Old Dominion University, Norfolk, Virginia 23529, USA}
\author{S.~Tkachenko}
\affiliation{University of Virginia, Charlottesville, Virginia 22901, USA}
\author{J.~Zhang}
\affiliation{University of Virginia, Charlottesville, Virginia 22901, USA}

\date{\today}

\begin{abstract}
Using data from the recent BONuS experiment at Jefferson Lab,
which utilized a novel spectator tagging technique to extract
the inclusive electron--free neutron scattering cross section,
we obtain the first direct observation of quark--hadron duality
in the neutron $F_2$ structure function.
The data are used to reconstruct the lowest few ($N=2$, 4 and 6)
moments of $F_2$ in the three prominent nucleon resonance regions,
as well as the moments integrated over the entire resonance region.
Comparison with moments computed from global parametrizations of
parton distribution functions suggest that quark--hadron duality holds
locally for the neutron in the second and third resonance regions
down to $Q^2 \approx 1$~GeV$^2$, with violations possibly up to
20\% observed in the first resonance region.
\end{abstract}


\maketitle

\section{Introduction}

Inclusive lepton scattering has for many decades been the most
important tool for probing the internal quark and gluon (or parton)
structure of nucleons and nuclei.  Structure functions extracted
from inclusive deep-inelastic scattering (DIS) experiments display
the central features of quantum chromodynamics (QCD) ---
asymptotic freedom at short distances ({\it via} structure function
scaling and its violation) and confinement at large distance scales
({\it via} the momentum dependence of parton distributions).

Since the late 1960s, DIS experiments have yielded an impressive 
data set that maps nucleon structure functions over several orders
of magnitude in the Bjorken scaling variable, $x$, and the squared
four-momentum transfer, $Q^2$.  These data, supplemented by cross
sections from hadronic collisions and other high-energy processes,
have enabled a detailed picture of the parton distribution functions
(PDFs) of the nucleon to emerge through global QCD analyses
(see Refs.~\cite{Jimenez13, Forte13} and references therein).

At lower energies, where nonperturbative quark--gluon interactions
are important and the inclusive lepton--nucleon cross section is
dominated by nucleon resonances, the structure functions reveal
another intriguing feature of QCD, namely, quark--hadron duality.
Here, the low energy cross section, when averaged over appropriate
energy intervals, is found to resemble the high energy result,
whose $Q^2$ dependence is described by perturbative QCD.
In this context, quark--hadron duality provides a unique
perspective on the relationship between confinement
and asymptotic freedom, and establishes a critical link between
the perturbative and nonperturbative regimes of QCD.

In the framework of QCD, quark--hadron duality can be formally
interpreted in terms of structure function moments \cite{DeRujula75}.
From the operator product expansion (OPE), the moments can be
expressed as a series in $1/Q^2$, with coefficients given by
matrix elements of local quark--gluon operators of a given twist
\cite{Wilson69}.
The leading (twist 2) term corresponds to scattering from a single
parton, while higher twist terms correspond to multi--quark
and quark--gluon interactions.
At low $Q^2$ the resonance region makes a significant contribution
to the structure function moments, so that here one might expect
a strong $Q^2$ dependence of the moments arising from the higher
twist terms in the OPE.
In practice, however, the similarity of the structure function moments
at low $Q^2$ and the moments extracted from high energy cross sections
suggests the dominance of the leading twist contribution.
The combined higher twist, multi-parton contributions appear to play
a relatively minor role down to scales of the order
$Q^2 \sim 1$~GeV$^2$.

This nontrivial relationship between the low-energy cross section
and its deep-inelastic counterpart was first observed by Bloom and
Gilman \cite{Bloom70, Bloom71} in the early DIS measurements that
were instrumental in establishing structure function scaling.
More recently, the availability of extensive, precise structure
function data from Jefferson Lab and elsewhere, over a wide range
of kinematics, has opened up the possibility for in-depth studies
of quark--hadron duality.
Duality has now been observed in the proton $F_2$ and $F_L$
structure functions \cite{Niculescu00, Liang, Malace09, Bianchi04,
Melnitchouk05, Monaghan13}, the $F_2$ structure function of nuclei
\cite{Niculescu06}, the spin-dependent $g_1$ structure functions
of the proton and $^3$He \cite{HERMES03, Bosted07, Solvignon08},
the individual helicity-1/2 and 3/2 virtual photoproduction cross
sections for the proton \cite{Malace11}, and in parity-violating 
electron--deuteron scattering \cite{PVDIS}.

To establish the dynamical origin of quark-hadron duality in the
nucleon requires one to also study the low-$Q^2$ structure of the
neutron.  Models based on four-quark higher twist contributions to
DIS suggest that duality in the proton could arise from accidental
cancellations between quark charges, which would not occur for the
neutron \cite{Brodsky00}.
Unfortunately, the absence of high-density free neutron targets
means that essentially all information on the structure functions
of the neutron has had to be derived from measurements on deuterium.
Typically, the deuterium data are corrected for Fermi smearing and
other nuclear effects \cite{Malace10, Arrington12, Osipenko06,
Weinstein11, Hen11, Arrington09, CJ11}, 
which introduces an element of model
dependence into the extraction procedure.
This is particularly problematic in the nucleon resonance region,
where Fermi motion effects leads to significant smearing of the
resonant structures.
The existence of duality in the neutron $F_2$ structure function
was suggested recently \cite{Malace10} in an analysis which used
an iterative deconvolution method \cite{Kahn08} to extract neutron
resonance spectra from inclusive proton and deuteron $F_2$ data
\cite{Malace09}.  A model independent confirmation of duality in the
neutron, however, was to date not possible.

Recently, a new experimental technique, based on spectator nucleon
tagging \cite{Fenker08}, has been used to extract the free neutron
$F_2$ structure function \cite{Baillie12}.  By detecting low-momentum
protons at backward angles in electron deuteron scattering, the BONuS
experiment at Jefferson Lab measured $F_2$ for the neutron in both
the resonance and DIS regions, with minimal uncertainty from nuclear
smearing and rescattering corrections \cite{Tkachenko14}.
In the present work, we use the BONuS data to quantitatively measure
for the first time the degree to which duality holds for the $F_2$
structure function of the free neutron.  Because the results reported
here use data from an experimentally--isolated neutron target,
one expects significantly reduced systematic uncertainties compared
with those in the model-dependent analysis of inclusive deuterium data
\cite{Malace10}.

For the theoretical analysis of duality we use the method of
truncated structure function moments \cite{Forte99, Forte01,
Piccione01, Kotlorz07}, which were applied to the resonance region
$F_2$ proton data by Psaker {\it et al.} \cite{Psaker08}.
Here, the $n$-th truncated moment of the $F_2$ structure function
is defined as
\begin{equation}
M_N(x_{\rm min},x_{\rm max},Q^2)
= \int_{x_{\rm min}}^{x_{\rm max}} dx\, x^{N-2} F_2(x,Q^2),
\label{eq:moments2}
\end{equation}
where the integration over $x$ is restricted to an interval
between $x_{\rm min}$ and $x_{\rm max}$.
This method avoids extrapolation of the integrand into poorly
mapped kinematic regions, and is particularly suited for the study
of duality where an $x$ interval can be defined by a resonance
width around an invariant mass $W^2 = M^2 + Q^2 (1-x)/x$,
where $M$ is the nucleon mass.
As the position of the resonance peak varies with $x$ for different
$Q^2$ values, the values for $x_{\rm min}$ and $x_{\rm max}$ evolve
to the appropriate invariant mass squared region.
For the BONuS data, we consider four ranges in $W^2$, corresponding
to the three prominent resonance regions as well as the combined
resonance region,
\begin{eqnarray}
\label{eq:Wregions}
&& 1.3 \leq W^2 \le 1.9~{\rm GeV}^2\ \ \ 
	\textrm{[1st (or $\Delta$) region]},	\nonumber\\
&& 1.9 \leq W^2 \le 2.5~{\rm GeV}^2\ \ \
	\textrm{[2nd region]},			\nonumber\\
&& 2.5 \leq W^2 \le 3.1~{\rm GeV}^2\ \ \
	\textrm{[3rd region]},			\\
&& 1.3 \leq W^2 \le 4.0~{\rm GeV}^2\ \ \
	\textrm{[total resonance]}.		\nonumber
\end{eqnarray}

After reviewing the BONuS experiment in Sec.~\ref{sec:experiment},
the results for several low truncated moments (corresponding to
$N=2$, 4 and 6) of the neutron $F_2$ structure function are
presented in Sec.~\ref{sec:truncated}.
The implications of the new data for local quark-hadron duality
and its violation are discussed by comparing with recent global
PDF parametrizations and previous model-dependent data analyses
(Sec.~\ref{ssec:n_duality}).
The isospin dependence of local duality is studied by comparing
the neutron moments with corresponding moments of the proton
$F_2$ structure function (Sec.~\ref{ssec:n_isospin}).
Finally, we summarize our results and conclusions in
Sec.~\ref{sec:conclusion}.

\section{The BONuS Experiment}
\label{sec:experiment}

The results reported here rely on a novel experimental technique aimed 
at eliminating or substantially reducing the theoretical uncertainties 
involved in extracting neutron data from nuclear targets.  The BONuS
(Barely Off--shell Nucleon Structure) experiment at Jefferson Lab
\cite{Fenker08, Baillie12, Tkachenko14} used a Radial Time Projection
Chamber (RTPC) to detect low momentum spectator protons produced in
electron--deuterium scattering in conjunction with electrons
detected with CLAS \cite{Mecking03} in Hall B.
The tagging technique essentially eliminates effects of Fermi smearing
\cite{FS88}, while the restriction to backward low-momentum spectator
protons minimizes final state interactions \cite{cdg04, Cosyn11,
Cosyn14} and ensures that the neutron is nearly on-shell
\cite{MSS97, Baillie12}.

The BONuS experiment ran in 2005 and acquired electron--deuteron
scattering data at two electron beam energies, $E=4.223$ and 5.262~GeV.
The RTPC consisted of three layers of gas electron multipliers
surrounding a thin, pressurized gas deuterium target, and was able
to detect protons with momenta as low as 70~MeV.
The experiment and data analysis are described in detail in
Ref.~\cite{Tkachenko14}.  Ratios of neutron to proton $F_2$ structure
functions and the absolute neutron $F_2$ structure function were
extracted over a wide kinematic range and for spectator proton momenta
between 70 and 100~MeV.  The total systematic uncertainty in the
neutron structure function extracted was 8.7\% \cite{Tkachenko14},
with an overall 10\% scale uncertainty.

\begin{figure}
\begin{center}
\includegraphics[scale=0.4]{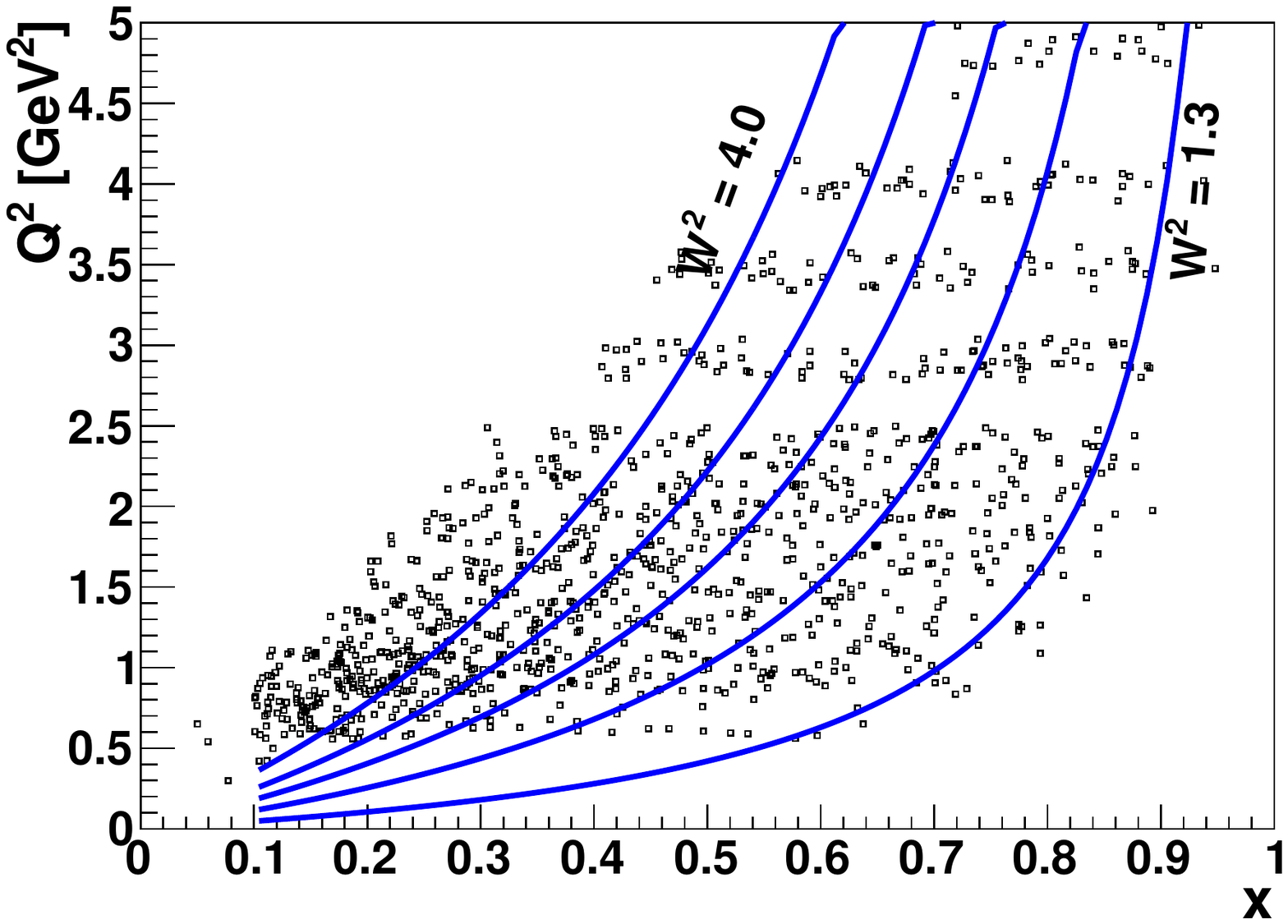}
\end{center}
\caption{Kinematic coverage of the BONuS data.
	The solid lines denote the fixed-$W^2$ thresholds for the
	four final state mass regions in Eq.~(\ref{eq:Wregions}),
	from $W^2=1.3$ to 4.0~GeV$^2$.}
\label{fig:bonus_kinematics}
\end{figure}

\begin{figure}
\begin{center}
\includegraphics[scale=0.43]{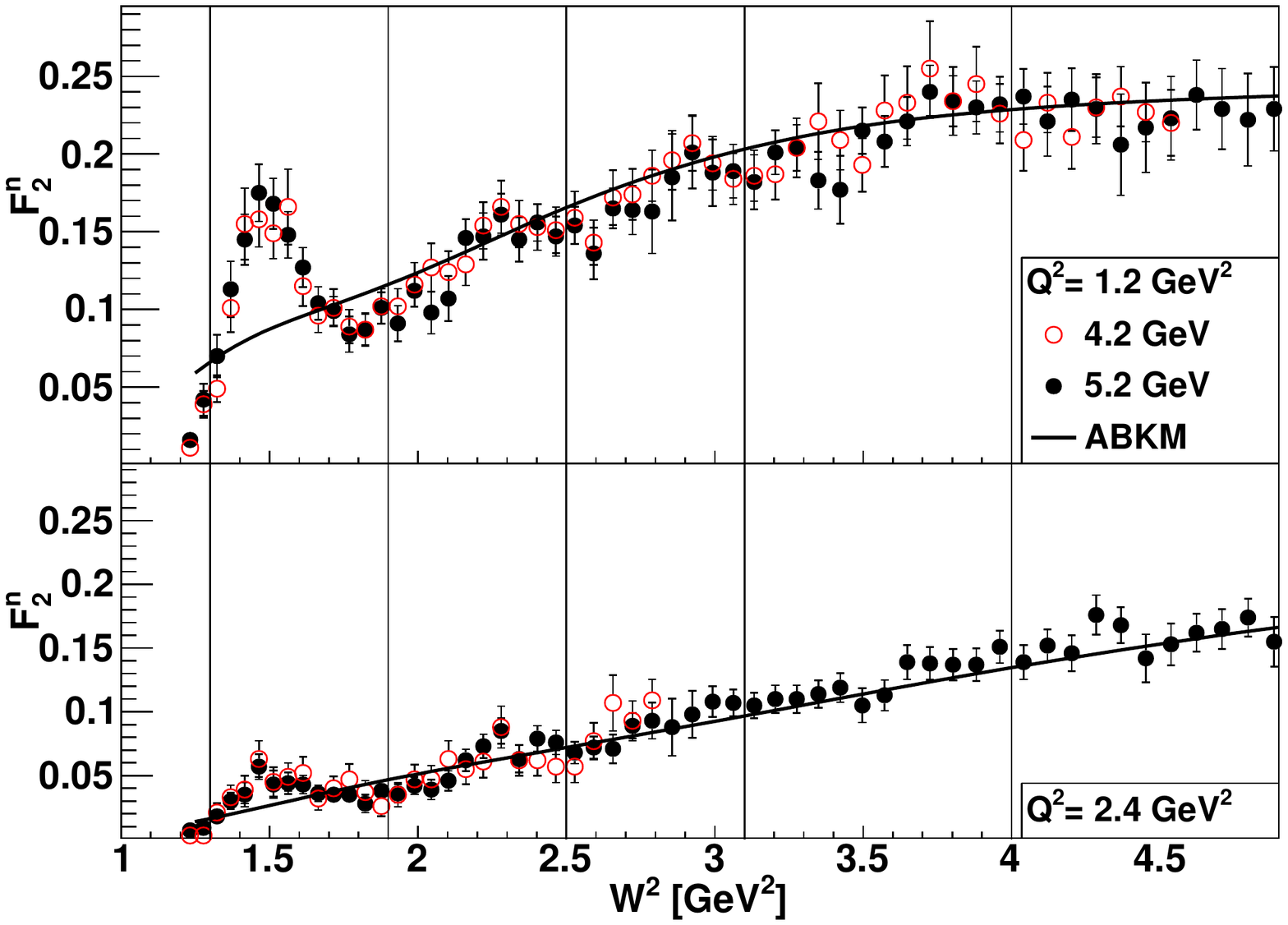}
\end{center}
\caption{Representative neutron $F_2^n$ structure function spectra
	from the BONuS experiment \cite{Tkachenko14} at
	$Q^2=1.2$~GeV$^2$ (top panel) and
	$Q^2=2.4$~GeV$^2$ (bottom panel).  The open (filled)
	circles represent data for a beam energy of $E=4.223$
	(5.262)~GeV.  The solid curve is computed from the ABKM
	global PDF parametrization \cite{ABKM} including higher
	twist effects and target mass corrections.}
\label{fig:spectra}
\end{figure}

The kinematic coverage, shown in Fig.~\ref{fig:bonus_kinematics}
(with the 4.223 and 5.262~GeV data combined), extends from the
threshold to the deep-inelastic region.  The curves in
Fig.~\ref{fig:bonus_kinematics} represent the fixed-$W^2$
thresholds for the four mass regions considered.
Typical neutron $F_2^n$ spectra are shown in Fig.~\ref{fig:spectra}
for $Q^2=1.2$ and 2.4~GeV$^2$, with the data restricted to spectator
proton angles greater than 100$\deg$ relative to the momentum
transfer, and proton momenta between 70 and 100~MeV.
The BONuS results are compared with the ABKM global fit \cite{ABKM}
to deep-inelastic and other high-energy scattering data, with the
inclusion of target mass corrections and higher twist effects.
The qualitative agreement between the parametrization and data
suggests evidence for quark--hadron duality, which we explore
more quantitatively in the following sections.

\section{Truncated Moments and Local Quark--Hadron Duality}
\label{sec:truncated}

Because the kinematic variables $Q^2$, $x$ and $W^2$ are not independent,
a range in $W^2$ at fixed $Q^2$ implies a corresponding range in $x$.
This makes possible a straightforward integration of the experimental
$F_2^n$ structure function data to obtain the truncated moments $M_n$
in Eq.~(\ref{eq:moments2}).  To minimize the model dependence, we
evaluate the integrals based solely on the experimentally measured
data points, without using any interpolating or extrapolating function.

\subsection{Truncated neutron moments}
\label{ssec:n_duality}

\begin{figure}
\begin{center}
\includegraphics[scale=0.43]{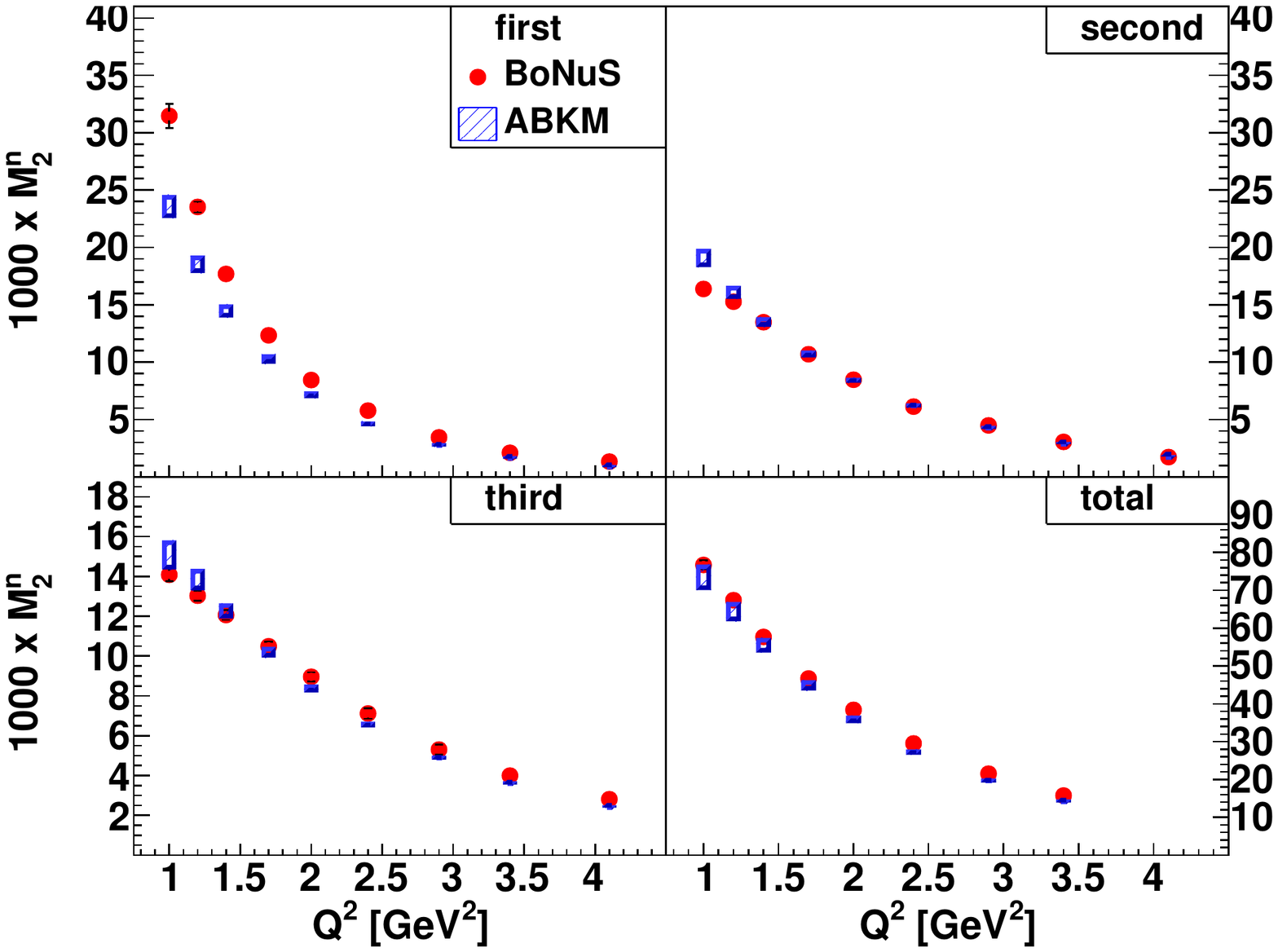}
\end{center}
\caption{Second ($N=2$) neutron truncated moments $M_2^n$ versus $Q^2$
	for the four resonance regions in Eq.~(\ref{eq:Wregions})
	[labeled as ``first'', ``second'', ``third'' and ``total''].
	The moments obtained from the BONuS data (filled circles)
	are compared with moments computed from the ABKM global PDF
	parametrization \cite{ABKM} including target mass and
	higher twist corrections (shaded rectangles).}
\label{fig:moments}
\end{figure}

The second ($N=2$) truncated neutron moments, $M_2^n$, obtained from
the BONuS data are shown in Fig.~\ref{fig:moments} as a function of
$Q^2$ for the four $W^2$ intervals defined in Eq.~(\ref{eq:Wregions}).
The numerical values for the moments are also listed in
Table~\ref{tab:moments}.
The quoted errors take into account the experimental statistical and
systematic uncertainties added in quadrature, but do not the include
the 10\% scale uncertainty.

The experimental moments are compared with the moments calculated
from the ABKM global PDF parametrization \cite{ABKM}, including
finite-$Q^2$ corrections from the target mass and an $x$-dependent
parameterization of the leading (${\cal O}(1/Q^2)$) higher twist
effects.
The latter are needed in order to obtain a more quantitative
description of duality in the low-$Q^2$ region, to which the
structure functions from the global fits (which are primarily
constrained by high energy data) are extrapolated.
%
%
The comparison shows generally very good agreement in the second
and third resonance regions, and in the total integrated $W^2$
interval, while the ABKM results underestimate the data somewhat
in the $\Delta$ resonance region.

The corresponding higher order truncated moments (for $N=4$ and $N=6$)
are listed in Tables~\ref{tab:moments4} and~\ref{tab:moments6},
respectively.  Comparison with the ABKM fit (not shown) reveals a
similar pattern as for the $N=2$ moments, although the deviation in
the lowest-$W$ interval is more pronounced, especially at low $Q^2$,
because of the greater weighting given to the high-$x$ region in
the higher moments.

\begin{table}[t]
\begin{tabular}{c|c|c|c|c} \hline
& \multicolumn{4}{c} {$M_2\, [\times 10^{-3}]$}   \\ \hline
$Q^2$ [GeV$^2$]& 1st & 2nd & 3rd & total \\ \hline
1.0 & 31.5 $\pm$ 1.1 & 16.4 $\pm$ 0.4 & 14.1 $\pm$ 0.3 & 76.7 $\pm$ 1.2 \\
1.2 & 23.5 $\pm$ 0.5 & 15.3 $\pm$ 0.3 & 13.0 $\pm$ 0.3 & 67.4 $\pm$ 0.6 \\
1.4 & 17.7 $\pm$ 0.4 & 13.5 $\pm$ 0.2 & 12.1 $\pm$ 0.3 & 57.7 $\pm$ 0.5 \\
1.7 & 12.3 $\pm$ 0.3 & 10.7 $\pm$ 0.2 & 10.5 $\pm$ 0.2 & 46.7 $\pm$ 0.5 \\
2.0 & 8.4 $\pm$ 0.2 & 8.5 $\pm$ 0.2 & 9.0 $\pm$ 0.2 & 38.4 $\pm$ 0.4 \\
2.4 & 5.8 $\pm$ 0.2 & 6.1 $\pm$ 0.1 & 7.1 $\pm$ 0.3 & 29.5 $\pm$ 0.4 \\
2.9 & 3.4 $\pm$ 0.1 & 4.5 $\pm$ 0.1 & 5.3 $\pm$ 0.3 & 21.5 $\pm$ 0.4 \\
3.4 & 2.1 $\pm$ 0.1 & 3.1 $\pm$ 0.1 & 4.0 $\pm$ 0.2 & 15.8 $\pm$ 0.3 \\
4.1 & 1.3 $\pm$ 0.1 & 1.7 $\pm$ 0.1 & 2.8 $\pm$ 0.1 &  --- \\ \hline
\end{tabular}
\caption{Second ($N=2$) truncated moments (in units of $10^{-3}$)
	of the neutron $F_2$ structure function from the BONuS data
	for the $W^2$ regions in Eq.~(\ref{eq:Wregions}).
	The errors are a quadrature sum	of statistical and systematic
	uncertainties, but do not include the overall 10\%
	normalization uncertainty.}
\label{tab:moments}
\end{table}
\begin{table}[h]
\resizebox{8.5 cm}{!}{
\begin {tabular}{c|c|c|c|c}\hline
& \multicolumn{4}{c} {$M_4\, [\times 10^{-3}]$}   \\ \hline
$Q^2$ [GeV$^2$]& 1st & 2nd & 3rd & total \\ \hline
1.0 & 11.58 $\pm$ 0.43 & 3.09 $\pm$ 0.08 & 1.69 $\pm$ 0.04 & 17.49 $\pm$ 0.44 \\
1.2 & 9.80 $\pm$ 0.21 & 3.51 $\pm$ 0.06 & 1.95 $\pm$ 0.04 & 16.78 $\pm$ 0.22 \\
1.4 & 8.11 $\pm$ 0.17 & 3.60 $\pm$ 0.06 & 2.17 $\pm$ 0.04 & 15.61 $\pm$ 0.19 \\
1.7 & 6.27 $\pm$ 0.14 & 3.40 $\pm$ 0.06 & 2.33 $\pm$ 0.05 & 14.01 $\pm$ 0.17 \\
2.0 & 4.67 $\pm$ 0.14 & 3.08 $\pm$ 0.06 & 2.36 $\pm$ 0.06 & 12.45 $\pm$ 0.17 \\
2.4 & 3.48 $\pm$ 0.11 & 2.54 $\pm$ 0.06 & 2.20 $\pm$ 0.08 & 10.59 $\pm$ 0.15 \\
2.9 & 2.22 $\pm$ 0.10 & 2.11 $\pm$ 0.07 & 1.93 $\pm$ 0.09 & 8.52 $\pm$ 0.16 \\
3.4 & 1.44 $\pm$ 0.09 & 1.58 $\pm$ 0.07 & 1.64 $\pm$ 0.08 & 6.72 $\pm$ 0.15 \\
4.1 & 0.95 $\pm$ 0.08 & 0.98 $\pm$ 0.07 & 1.29 $\pm$ 0.06 &  --- \\ \hline
\end{tabular}}
\caption{As in Table~\ref{tab:moments}, but for the $N=4$ moment.}
\label{tab:moments4}
\end{table}
\begin{table}[b]
\resizebox{8.5 cm}{!}{
\begin {tabular}{c|c|c|c|c}\hline
& \multicolumn{4}{c} {$M_6\, [\times 10^{-3}]$}   \\ \hline
$Q^2$ [GeV$^2$]& 1st & 2nd & 3rd & total \\ \hline
1.0 & 4.39 $\pm$ 0.18 & 0.60 $\pm$ 0.01 & 0.20 $\pm$ 0.01 & 5.28 $\pm$ 0.18 \\
1.2 & 4.19 $\pm$ 0.10 & 0.82 $\pm$ 0.01 & 0.30 $\pm$ 0.01 & 5.45 $\pm$ 0.10 \\
1.4 & 3.79 $\pm$ 0.09 & 0.98 $\pm$ 0.02 & 0.39 $\pm$ 0.01 & 5.38 $\pm$ 0.09 \\
1.7 & 3.24 $\pm$ 0.08 & 1.09 $\pm$ 0.02 & 0.52 $\pm$ 0.01 & 5.17 $\pm$ 0.08 \\
2.0 & 2.62 $\pm$ 0.08 & 1.13 $\pm$ 0.02 & 0.62 $\pm$ 0.02 & 4.82 $\pm$ 0.09 \\
2.4 & 2.12 $\pm$ 0.07 & 1.06 $\pm$ 0.02 & 0.68 $\pm$ 0.02 & 4.41 $\pm$ 0.08 \\
2.9 & 1.45 $\pm$ 0.07 & 1.00 $\pm$ 0.03 & 0.71 $\pm$ 0.03 & 3.77 $\pm$ 0.08 \\
3.4 & 0.99 $\pm$ 0.07 & 0.82 $\pm$ 0.04 & 0.67 $\pm$ 0.03 & 3.14 $\pm$ 0.09 \\
4.1 & 0.68 $\pm$ 0.06 & 0.56 $\pm$ 0.04 & 0.60 $\pm$ 0.03 & ---  \\ \hline
\end{tabular}}
\caption{As in Table~\ref{tab:moments}, but for the $N=6$ moment.}
\label{tab:moments6}
\end{table}

Note that while early phenomenological analyses of quark--hadron
duality typically compared resonance region data at low $Q^2$ with
scaling functions extrapolated from fits to high-$W$ cross sections
\cite{Bloom70, Bloom71}, more recent quantitative analyses
\cite{Malace09, Malace10} have emphasized the need to take
into account the $Q^2$ dependence in the high-$W$ data,
including both leading and higher twist contributions.
This is especially important in the high-$x$ region, where the
separation between the leading and higher twists is more model
dependent due to the absence of high-$Q^2$ measurements, and
comparison of resonance region data with the total extrapolated
structure functions reveals an enhanced persistence of duality
down to lower values of $Q^2$.

\begin{figure}
\begin{center}
\includegraphics[scale=0.42]{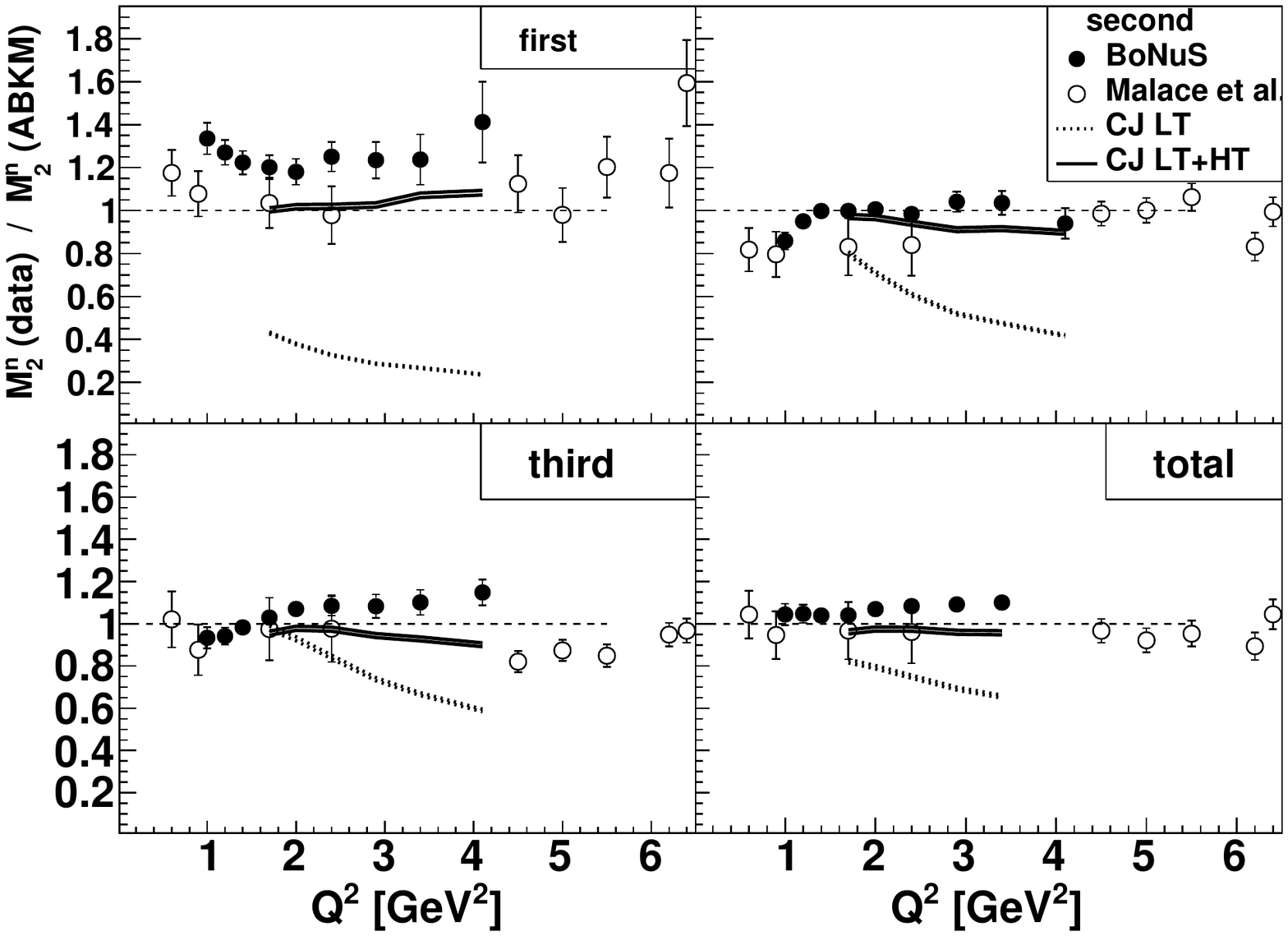}
\end{center}
\caption{Ratios of truncated moments of the neutron $F_2$ structure
	function from the BONuS data to those computed from the ABKM
	global PDF parametrization \cite{ABKM} including finite-$Q^2$
	effects	(filled circles) as a function of $Q^2$ for the four
	$W^2$ intervals in Eq.~(\ref{eq:Wregions}).
	The empirical moments are compared with the results of the
	model-dependent analysis of inclusive DIS data \cite{Malace10}
	(open circles), and with ratios computed from the CJ12
	distributions \cite{CJ12}, with leading twist only
	(dotted lines) and including finite-$Q^2$ effects
	(solid lines).
	All ratios are taken relative to the ABKM moments.}
\label{fig:data_theory_m2}
\end{figure}

To study local quark--hadron duality in detail, we form ratios of
the truncated moments of $F_2^n$ obtained from the BONuS data to the
moments computed from the ABKM reference structure function \cite{ABKM},
over the same range of $x$.  The ratios for the $M_2^n$ moments are
shown in Fig.~\ref{fig:data_theory_m2} as a function of $Q^2$ for
the four invariant mass regions in Eq.~(\ref{eq:Wregions}).
The ratios for the second, third and total resonance regions are
close to unity, to within $\sim 10\%$ over nearly the entire range
of $Q^2=1-4$~GeV$^2$, and exhibit weak scale dependence.
This points to a dramatic confirmation of the validity of local
duality for the neutron in these regions.
In the first resonance region, the $\Delta$ resonance is
$\sim 20\%-30\%$ larger than the PDF-based fit, but still displays
a similar $Q^2$ behavior.
This could be interpreted as either a stronger violation of local
duality in the $\Delta$ region, which may be expected at lower $W$,
or possibly underestimated strength of the ABKM parametrization
in the large-$x$ regime, to which this $W$ region corresponds.
While this is difficult to disentangle experimentally, duality is
expected to hold to better accuracy for integrals obtained over
regions containing multiple final states.

The confirmation of the approximate validity of duality in $F_2^n$
from the BONuS data disfavors the suggestion \cite{Brodsky00} that
duality occurs in the proton because of accidental cancellations of
quark charges associated with higher twist, four-quark operators,
and disagrees with the prediction that duality should therefore
not be seen in the neutron.
This conclusion was also reached in the model-dependent analysis
by Malace {\it et al.} \cite{Malace10}, who studied duality in the
neutron by extracting the $F_2^n$ structure function from inclusive
DIS data using phenomenological deuteron wave functions and an
iterative deconvolution procedure \cite{Kahn08}.
Overall, the BONuS data are in good agreement with the earlier
results \cite{Malace10}, within the experimental uncertainties,
although they appear to lie systematically higher in the $\Delta$
region.
This may be associated with the nuclear corrections in the deuteron,
which are subject to greater uncertainties at the largest $x$
(smallest $W$) values, or a systematic bias of the subtraction
method in relation to the various theoretical assumptions and
models \cite{Arrington12}.

The relevance of large-$x$ uncertainties and finite-$Q^2$ corrections
in global PDF fits is illustrated in Fig.~\ref{fig:data_theory_m2},
where the experimental and computed ABKM moments are also compared
with the moments calculated from the CTEQ--Jefferson Lab (CJ) global
PDF parametrization \cite{CJ12} with and without higher twist
corrections.
While the ratio of the ABKM and CJ moments is close to unity over the
entire range of $Q^2$ considered when finite-$Q^2$ effects are included,
the deviation from unity of the ratio computed from only the leading
twist components of the CJ fit can be up to $30\%-40\%$ for the
integrated resonance region, and up to twice as much for the
$\Delta$ region.
This suggests an important role played by the finite-$Q^2$ corrections
to the scaling functions in effecting the cancellations between the
individual resonance regions necessary for the realization of
quark-hadron duality \cite{Isgur01, Close01, Close03}.

However, even incorporating finite-$Q^2$ corrections, global PDF
fits can differ significantly in the large-$x$ (low-$W$) regime.
Because of the $Q^2$ and $W^2$ cuts applied to the global data set,
PDFs in the largest-$x$ regions relevant for this analysis are
essentially unconstrained, resulting in large uncertainties in
the extrapolated $x \to 1$ behavior \cite{CJ10}.
%

\subsection{Isospin dependence}
\label{ssec:n_isospin}

\begin{figure}
\begin{center}
\includegraphics[scale=0.43]{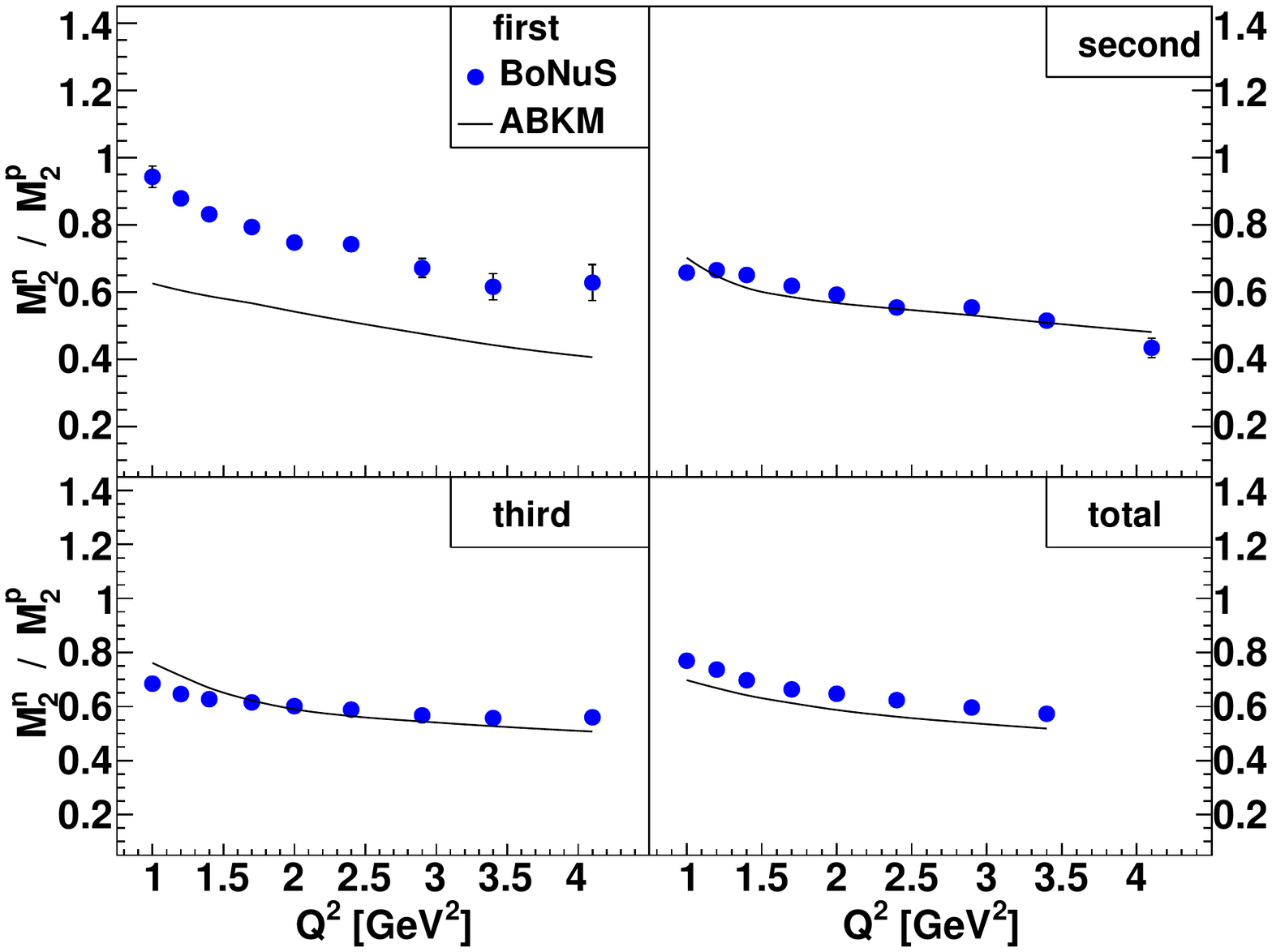}
\end{center}
\caption{Ratios $M_2^n/M_2^p$ of neutron to proton truncated moments
	of the $F_2$ structure function versus $Q^2$, for the four
	$W$ regions in Eq.~(\ref{eq:Wregions}).  The BONuS results
	(filled circles) are compared with the moments computed from
	the ABKM global PDF parametrization including target mass
	and higher twist (solid lines) corrections.
	In both cases the proton moments are evaluated from the
	same ABKM fit \cite{ABKM}.}
\label{fig:data_f2nf2p}
\end{figure}

The stronger violation of local duality in the $\Delta$ region is
also evident in the ratio of neutron to proton truncated moments,
shown in Fig.~\ref{fig:data_f2nf2p} compared with the reference
ABKM parametrization \cite{ABKM} that was used to compute both
the proton and neutron moments.  To obtain the empirical proton
truncated moments in the resonance region, the Christy-Bosted
global fit \cite{Christy10} was utilized.
(Duality in the proton structure function moments themselves was
studied in detail in previous analyses \cite{Malace09}, and
generally confirmed at the $10\%-15\%$ level for the $N=2$ moment
when integrated over the entire resonance region.)
%

The significant duality violation in the neutron/proton ratio observed
in the $\Delta$ region can be understood from the isovector nature
of the $\Delta$-isobar and the relatively small nonresonant
background on which it sits.
In the limit of exact isospin symmetry, the transitions from a ground
state nucleon to an isospin-3/2 resonance would be identical for
protons and neutrons.  Nonresonant background and isospin symmetry
breaking contributions give rise to differences between proton and
neutron moments, but these are typically very small in the $\Delta$
region.
In contrast, the proton and neutron deep-inelastic structure functions
(either leading twist only or with higher twist corrections) in the 
$\Delta$ region are expected to be quite different, since at large $x$
the neutron structure function is strongly suppressed relative to the
proton, $F_2^n \ll F_2^p$ \cite{Close88, Meln96}.
The fact that the experimental $M_2^n/M_2^p$ ratio in the high-$x$
region lies somewhat higher than the DIS parametrization (even more
pronounced than in Ref.~\cite{Malace10}) is therefore consistent
with these expectations.

A similar comparison of the neutron to proton moments in the
second and third resonance regions in Fig.~\ref{fig:data_f2nf2p}
shows significantly better agreement with the DIS parametrization.
Based on simple quark models and assuming magnetic coupling dominance,
one would expect the resonance contribution to the neutron moments
to underestimate the DIS moment in the second resonance region.
This is due to the small couplings to octet spin-1/2 states.
In contrast, according to Refs.~\cite{Close01, Close03, Close09}
the proton moments would overestimate the DIS results in the
second and third $W$ intervals in Eq.~(\ref{eq:Wregions}).
While there was some evidence for such a pattern from the earlier,
model-dependent analysis of inclusive data \cite{Malace10}, there
is no indication from the BONuS results of a suppression in the
second resonance region.  The slightly larger overall magnitude
of the neutron moments compared with Ref.~\cite{Malace10} brings
the present results into excellent agreement with the DIS moments
in the second region, with a small enhancement in the third region.
The corresponding enhancement of the proton data in the
third resonance region relative to the ABKM fit \cite{Niculescu00,
Malace09} then results in essentially no deviation of the neutron
to proton ratio here, as illustrated in Fig.~\ref{fig:data_f2nf2p}.
Finally, for the total integrated region between threshold and
$W = 2$~GeV, the empirical $M_2^n/M_2^p$ ratio is slightly above
the DIS result mostly because of the large enhancement of the
data in the $\Delta$ region.

\section{Conclusion}
\label{sec:conclusion}

In this work we have investigated local quark--hadron duality in the
neutron structure function based on data from the BONuS experiment
at Jefferson Lab \cite{Baillie12, Tkachenko14}, which used a novel
experimental technique to create an effective neutron target by
tagging low momentum spectator protons in electron-deuterium
scattering.  The spectator tagging technique provides smaller
systematic uncertainties compared with the traditional method of
subtracting smeared hydrogen data and from inclusive deuterium
structure functions, using model assumptions for the nuclear
corrections.

We have evaluated the $N=2$, 4 and 6 truncated moments of the neutron
$F_2^n$ structure function for the three standard nucleon resonance
regions and the total integrated resonance region up to $W = 2$~GeV,
over the range $Q^2 = 1.0$ to 4.1~GeV$^2$.
Comparison of the experimental moments with moments computed from
global parametrizations of PDFs fitted to deep-inelastic and other
high energy scattering data, as well as with the corresponding
truncated moments for the proton, reveals a dramatic confirmation
of local duality for the neutron in the second, third and total
resonance regions to better than 10\% for the lowest moment.
The stronger ($\sim 20\%-30\%$) violation of duality in the
$\Delta$ region is consistent with the expectations based on
isospin symmetry for the isovector transition amplitudes and the
behavior of the $F_2^n/F_2^p$ ratio at large $x$ \cite{CJ12, Meln96}.

The confirmation of local duality in the neutron disfavors the model
\cite{Brodsky00} in which duality in the proton arises through
accidental cancellations of quark charges associated with higher
twist, four-quark operators, which would predict strong duality
violations in the neutron.  Rather, it suggests a dynamical origin
of duality in which cancellations among nucleon resonances produce
a higher degree of duality over the entire resonance region, with
stronger violations locally \cite{Close01, Close03, Close09}.
On the other hand, detailed comparisons between the empirical
truncated moments and DIS parametrizations in the individual
resonance regions suggest a pattern of duality violation that is
more involved than that predicted by simple spin-flavor symmetric
quark models with magnetic coupling dominance.

Our results also confirm and refine the findings of earlier
model-dependent studies \cite{Malace10} of duality in the neutron
in which the neutron structure was extracted from inclusive proton
and deuteron data assuming a model for the nuclear corrections
and an iterative deconvolution procedure \cite{Kahn08}.
In particular, the BONuS moments are found to lie slightly higher
than the earlier results, especially in the $\Delta$ region,
but with a similar $Q^2$ dependence.

In the future, the spectator tagging technique will be used at
Jefferson Lab with an 11~GeV electron beam to extend the kinematical
coverage of $F_2^n$ measurements to higher values of $x$ and $Q^2$
\cite{BONuS12}.  As well as providing more stringent constraints on
the leading twist PDFs in the limit $x \to 1$, the new data will
allow more definitive tests of local quark-hadron duality for the
neutron over a greater range of $Q^2$.

\acknowledgments

This work was supported by the United States Department of Energy (DOE)
Contract No.~DE-AC05-06OR23177, under which Jefferson Science Associates,
LLC operates Jefferson Lab. 
The JMU group was supported by the National Science Foundation (NSF) 
under Grant No.~PHY--1307196. 
S.K., J.A., S.T., and K.G. acknowledge support from the DOE under grants
DE-FG02-96ER40960,
DE-AC02-06CH11357,
DE-FG02-97ER41025, and
DE-FG02-96ER41003, respectively.


\end{document}